 \documentclass[reprint, amsmath, amssymb, aps, prb]{revtex4-1}

\usepackage[]{graphicx}      % Include figure files
\usepackage{bm}             	% bold math
\usepackage{times}			% Times font
\usepackage{tabularx}
\usepackage{color}
\usepackage{dcolumn}		% Align table columns on decimal point
\usepackage{amsmath}
\usepackage{amssymb}
\usepackage{amsfonts}
\usepackage{times}
\usepackage{tabularx}
\usepackage{xcolor,colortbl}

\begin{document}

\title{Material developments and domain wall based nanosecond-scale switching process in perpendicularly magnetized STT-MRAM cells}

\author{Thibaut Devolder}
\email{thibaut.devolder@u-psud.fr}
\author{Joo-Von Kim}
\affiliation{Centre de Nanosciences et de Nanotechnologies, CNRS, Univ. Paris-Sud, Universit\'e Paris-Saclay, C2N-Orsay, 91405 Orsay cedex, France}
\author{J. Swerts}
\author{S. Couet}
\author{S. Rao}
\author{W. Kim}
\author{S. Mertens}
\author{G. Kar} 
\affiliation{IMEC, Kapeldreef 75, B-3001 Leuven, Belgium}
\author{V. Nikitin}
\affiliation{SAMSUNG Electronics Corporation, 601 McCarthy Blvd Milpitas, CA 95035, USA}

\begin{abstract}
We investigate the Gilbert damping and the magnetization switching of perpendicularly magnetized FeCoB-based free layers embedded in magnetic tunnel junctions adequate for spin-torque operated magnetic memories. We first study the influence of the boron content in MgO / FeCoB /Ta systems alloys on their Gilbert damping parameter after crystallization annealing. Increasing the boron content from 20 to 30\% increases the crystallization temperature, thereby postponing the onset of elemental diffusion within the free layer. This reduction of the interdiffusion of the Ta atoms helps maintaining the Gilbert damping at a low level of 0.009 without any penalty on the anisotropy and the magneto-transport properties up to the 400$^\circ$C annealing required in CMOS back-end of line processing. In addition, we show that dual MgO free layers of composition MgO/FeCoB/Ta/FeCoB/MgO have a substantially lower damping than their MgO/FeCoB/Ta counterparts, reaching damping parameters as low as 0.0039 for a 3 \r{A} thick Tantalum spacer. This confirms that the dominant channel of damping is the presence of Ta impurities within the FeCoB alloy. 
On optimized tunnel junctions, we then study the duration of the switching events induced by spin-transfer-torque. We focus on the sub-threshold thermally activated switching in optimal applied field conditions. From the electrical signatures of the switching, we infer that once the nucleation has occurred, the reversal proceeds by a domain wall sweeping though the device at a few 10 m/s. The smaller the device, the faster its switching. We present an analytical model to account for our findings. The domain wall velocity is predicted to scale linearly with the current for devices much larger than the wall width. The wall velocity depends on the Bloch domain wall width, such that the devices with the lowest exchange stiffness will be the ones that host the domain walls with the slowest mobilities.
\end{abstract}

\maketitle

\section{Introduction}

Tunnel magnetoresistance (TMR) and spin transfer torque (STT) -- the fact that spin-polarized currents manipulate the magnetization of nanoscale magnets and in particular magnetic tunnel junction (MTJ) nanopillars -- are the basic phenomena underpinning an emerging technology called Spin-Transfer-Torque Magnetic Random Access Memory (STT-MRAM) \cite{Khvalkovskiy_basic_2013}, which combines high endurance, low power requirement \cite{chen_advances_2010, thomas_perpendicular_2014}, CMOS back-end-of-line (BEOL) compatibility \cite{swerts_beol_2015} and potentially large capacity \cite{sato_properties_2014}. 

The core of an STT-MRAM stack is a magnetic tunnel junction composed \cite{worledge_spin_2011} of an FeCoB/MgO/FeCoB central block. One of the FeCoB layer is pinned to a high anisotropy synthetic ferrimagnet to create a fixed reference layer (RL) system while the second FeCoB acts as a free layer (FL). Historically, the FL is capped with (or deposited on) an amorphous metal such as Ta \cite{ikeda_perpendicular-anisotropy_2010, swerts_beol_2015} and more recently capped with a second MgO layer to benefit from a second interface anisotropy \cite{ikeda_perpendicular-anisotropy_2010, sato_perpendicular-anisotropy_2012, konoto_effect_2013} in the so-called 'dual MgO' configuration. So far, it is unclear whether this benefit of anisotropy can be obtained without sacrificing the other important properties of the free layer, in particular the Gilbert damping. 

In this paper, we will first tailor the Boron content inside the FeCoB alloy to improve the properties of Ta / FeCoB / MgO 'single MgO' free layers and their resilience to thermal annealing. The idea is to postpone the FeCoB crystallization till the very last stage of the BEOL annealing. Indeed maintaining the amorphous state of FeCoB allows to minimize the interdiffusion of materials --in our case: tantalum-- within the stack. This interdiffusion is otherwise detrimental to the Gilbert damping. \\
We then turn to dual MgO systems comprising a Ta spacer layer in the midst of the FL. This spacer is empirically needed to allow proper crystallization and to effectively get perpendicular magnetic anisotropy (PMA) \cite{sato_perpendicular-anisotropy_2012, sinha_influence_2015, gottwald_scalable_2015, chang_effect_2013, sabino_influence_2014, kim_ultrathin_2015}. Unfortunately, the presence of heavy elements inside the FeCoB free layer is expected to alter its damping and to induce some loss of magnetic moment usually referred as the formation of magnetically dead layers. We study to what extend the Ta spacer in the dual MgO free layers affects the damping and how this damping compares with the one that can be obtained with single MgO free layers. Once optimized, damping factors as low as 0.0039 can be obtained a dual MgO free layer. 
 
	Besides the material issues, the success of STT-MRAM also relies on the capacity to engineer devices in accordance with industry roadmaps concerning speed and miniaturization. 	To achieve fast switching and design devices accordingly optimized, one needs to elucidate the physical mechanism by which the magnetization switches by STT. Several categories of switching modes -- macrospin \cite{sun_spin-current_2000}, domain-wall based \cite{bernstein_nonuniform_2011}, based on sub-volume nucleation \cite{sun_effect_2011} or based on the spin-wave amplification \cite{munira_calculation_2015} -- have been proposed, but single-shot time-resolved experimental characterization of the switching path are still scarce \cite{devolder_time-resolved_2016, devolder_size_2016, hahn_time-resolved_2016}. 
	Here we study the nanosecond-scale spin-torque-induced switching in perpendicularly magnetized tunnel junctions with sizes from 50 to 300 nm. Our time-resolved experiments argue for a reversal that happens by the motion of a single domain wall, which sweeps through the sample at a velocity set by the applied voltage. As a result, the switching duration is proportional to the device length. We model our finding assuming a single wall moving in a uniform material as a result of spin torque. The wall moves with a time-averaged velocity that scales with the product of the wall width and the ferromagnetic resonance linewidth, such that the devices with the lowest nucleation current densities will be the ones that host the domain walls with the lowest mobilities.

The paper is split in first a material science part, followed by a study of the magnetization reversal dynamics. After  a description of the samples and the caracterization methods, section \ref{BC} describes how to choose the optimal Boron content in an FeCoB-based free layer for STT-MRAM applications. Section \ref{SD} discusses the benefits of 'dual MgO' free layers when compared to 'single MgO' free layers. Moving to the magnetization switching section, the part \ref{methods} gathers the description of the main properties of the samples and the experimental methods used to characterize the STT-induced switching speed. Section \ref{results} describes the electrical signatures of the switching mechanism at the nanosecond scale. The latter is modeled in section \ref{model} in an analytical framework meant to clarify the factors that govern the switching speed when the reversal involves domain wall motion.

%%%%%%%%%%%%%%%%%%%%%%%%%%%%%%%%%%%%%%%%%%%%%%%%%%%%%%%%%%%%%%
\section{Advanced free layer designs}
\subsection{Model systems under investigation} %%%%%%%%%%%%%%%%%%%%%%%%%%%%%%%%%%%%%%%%%%%%%%%%%%%%%%%%%%%%%%
Our objective is to study advanced free layer designs in full STT-MRAM stacks. The stacks were deposited by physical vapor deposition in a Canon-Anelva EC7800 300 mm cluster tool. The MgO tunnel barriers were deposited by RF-magnetron sputtering. In dual MgO systems, the top MgO layer was fabricated by oxidation of a thin metallic Mg film. All stacks were post-deposition annealed in a TEL-MSL MRT5000  batch furnace in a 1 T perpendicular magnetic field for 30 minutes. Further annealing at 400$^\circ$C were done in a rapid thermal annealing furnace in a $\textrm{N}_2$ atmosphere for a period of 10 minutes. 

We will focus on several kinds of free layers embodied in state-of-the art bottom-pinned Magnetic Tunnel Junctions (MTJ) with various reference systems comprising either [Co/Ni] and [Co/Pt] based hard layers \cite{devolder_evolution_2016, devolder_annealing_2017}. Although we shall focus here on FLs deposited on [Co/Ni] based synthetic antiferromagnet (SAF) reference layers, we have conducted the free layer development also on [Co/Pt] based reference layers. While specific reference layer optimization leads to slightly different baseline TMR properties, we have found that the free layer performances were not impacted provided the SAF structure is stable with the concerned heat treatment (not shown). 

The first category of samples are the so-called 'single-MgO' free layers. We shall focus on samples with a free layer consists of a 1.4 nm thick Fe$_{60}$Co$_{20}$B$_{20}$ or a 1.6 nm thick Fe$_{52.5}$Co$_{17.5}$B$_{30}$ layer sandwiched between the MgO tunnel oxide and a Ta (2 nm) metal cap. Note that these so-called "boron 20\%" and "boron 30\%" samples have different boron contents but have the same number of Fe+Co atoms. A sacrificial \cite{swerts_beol_2015} Mg layer is deposed before the Ta cap to avoid Ta and FeCoB mixing during the deposition, and avoid the otherwise resulting formation of a dead layer. The Mg thickness is calibrated so that the Mg is fully sputtered away upon cap deposition. This advanced capping method has proven to provide improved TMR ratios and lower RA products thanks to an improved surface roughness and a higher magnetic moment \cite{swerts_beol_2015}.

The second category of free layers are the so-called 'dual MgO' free layers in which the FeCoB layer is sandwiched by the MgO tunnel oxide and an MgO cap which concur to improve the magnetic anisotropy. The exact free layer compositions are MgO (1.0 nm) /  Fe$_{60}$Co$_{20}$B$_{20}$ (1.1 nm) / spacer /  Fe$_{60}$Co$_{20}$B$_{20}$ (0.9 nm) / MgO (0.5 nm). We study shall two spacers: a Mg/Ta(3 \AA) spacer and a Mg/Ta(4 \AA) spacer, both comprising a sacrificial Mg layer.

%%%%%%%%%%%%%%%%%%%%%%%%%%%%%%%%%%%%%%%%%%%%%%%%%%%%%%%%%%%%%%
\subsection{Experimental methods used for material quality assessment} %%%%%%%%%%%%%%%%%%%%%%%%%%%%%%%%%%%%%%%%%%%%%%%%%%%%%%%%%%%%%%
\begin{figure}
\includegraphics[width=8.6 cm]{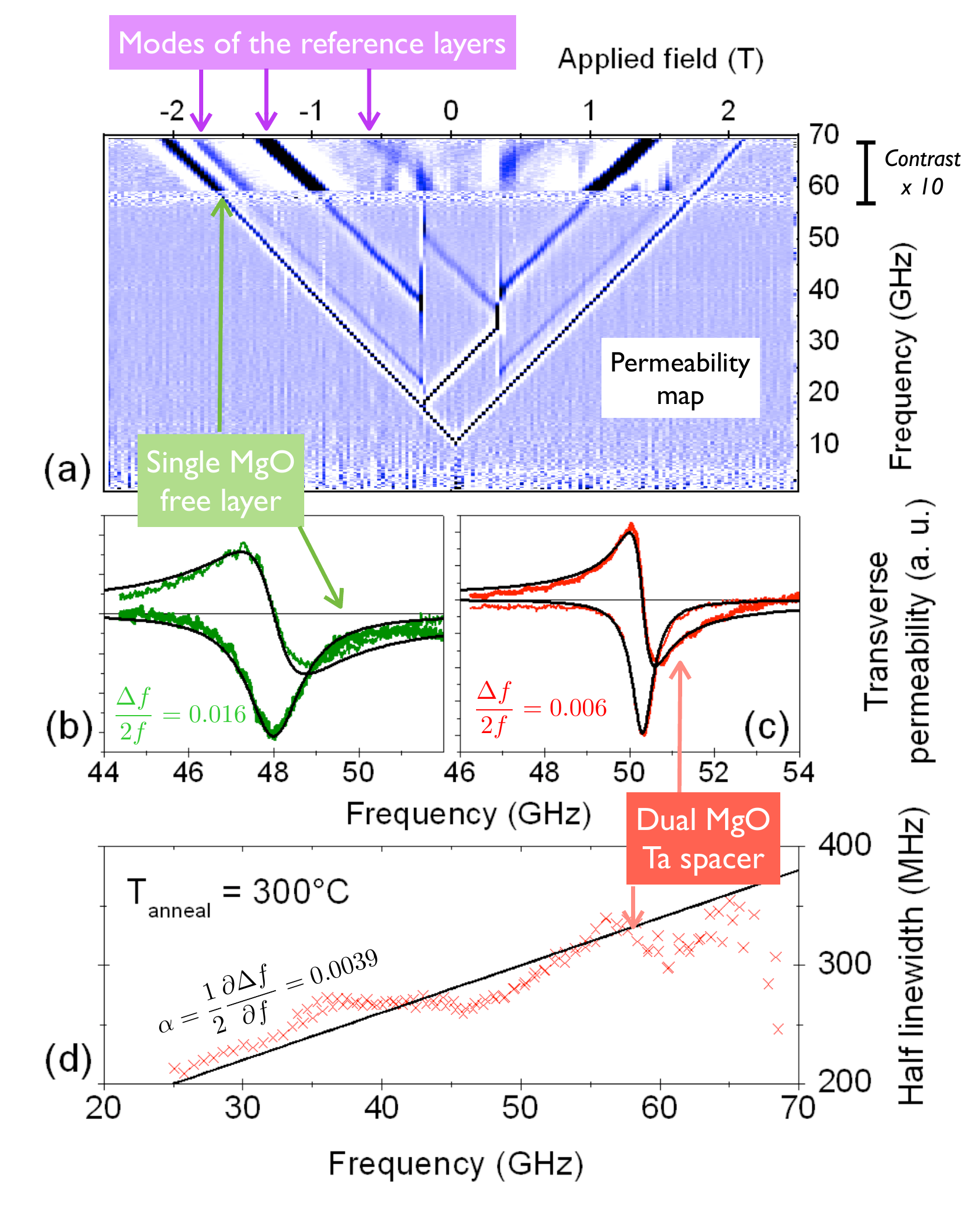}
\caption{(Color online). Examples of MTJ dynamical properties to illustrate the method of analysis. (a) Microwave permeability versus increasing out-of-plane field and frequency for an MTJ with a single MgO free layer %AE147719D06A
after an annealing of 300$^\circ$C. Note that the scale of the permeability was increased by a factor of 10 above 58 GHz for a better contrast. The apparent vertical bars are the eigenmode frequency jumps at the different switching fields of the MTJ. (b) Real and imaginary parts of the experimental (symbols) and modeled (lines) permeability for an out-of-plane field of 1.54 T for the same MTJ. The model is for an effective linewidth $\Delta f / (2f)= 0.016$, which includes both the Gilbert damping and a contribution from the sample inhomogeneity. (c) Same but for a dual MgO free layer based on a 3 \r{A} Ta spacer, modeled with $\Delta f / (2f)= 0.006$. (d) Cross symbols: FMR half frequency linewidth versus FMR frequency for a dual MgO free layer based on a 3 \r{A}  Ta spacer. The line is a guide to the eye corresponding to a Gilbert damping of 0.0039.}
\label{fmr}
\end{figure}

We studied our samples by current-in-plane tunneling (CIPT), vibrating sample magnetometry (VSM) and Vector Network Ferromagnetic resonance (VNA-FMR) \cite{bilzer_vector_2007} in out-of-plane applied fields. CIPT was performed to extract the tunnel magneto-resistance (TMR) and the resistance-area product (RA) of the junction. VSM measurements of the free layer minor loops have been used to extract the areal moments. We then use VNA-FMR to identify selectively the properties of each subsystem. Our experimental method is explained in Fig.~\ref{fmr}, which gathers some VNAFMR spectra recorded on optimized MTJs. The first panel records the permeability of a single MgO MTJ in the \{field-frequency\} parameter space. We systematically investigated a sufficiently large parameter space to detect 4 different modes whose spectral characters can be used to index them \cite{devolder_evolution_2016}. Three of the modes belong to the reference system that comprises 3 magnetic blocks coupled by interlayer exchange coupling through Ru and Ta spacers as usually done \cite{devolder_evolution_2016, devolder_annealing_2017}; the properties of these 3 modes are independent from the nature of the free layer. While we are not presently interested in analyzing the modes of the fixed system -- thorough analyses can be found in ref. \cite{devolder_evolution_2016, devolder_annealing_2017} -- we emphasize that it is necessary to detect all modes to unambiguously identify the one belonging to the free layer, in order to study it separately. The free layer modes are the ones having V-shaped frequency versus field curves [Fig.~\ref{fmr}(a)], whose slope changes at the free layer coercivity. in each sample, the free layer modes showed an asymmetric Lorentzian dispersion for the real part of the permeability and a symmetric Lorentzian dispersion for the imaginary part [see the examples Fig.~\ref{fmr}(b, c)]. As we found no signature of the two-layer nature of the dual MgO free layers, we modeled each free layer as a \textit{single} macrospin, disregarding whether it was a single MgO or a dual MgO free layer.

%%%%%%%%%%%%%%%%%%%%%%%%%%%%%%%%%%%%%%%%%%%%%%%%%%%%%%%%%%%%%%
%	Figure
%%%%%%%%%%%%%%%%%%%%%%%%%%%%%%%%%%%%%%%%%%%%%%%%%%%%%%%%%%%%%%

FMR frequency versus field fits [see one example in fig.~\ref{BoronContent}(c)] were used to get the effective anisotropy fields $H_k-M_s$ of all free layers \cite{devolder_damping_2013}. The curve slopes are $\gamma_0$, where $\gamma_0= 230$ kHz.m/A \textcolor{black}{is the gyromagnetic factor $\gamma$ multiplied by the vacuum permeability $\mu_0$. It was consistent with a spectroscopic splitting Land\'e factor of $g \approx 2.08$}. Damping analysis was conducted as follows: the free layer composition can yield noticeable differences in the FMR linewidths [see for instance Fig.~\ref{fmr}(b) and (c)]. To understand these differences, we systematically separated the Gilbert damping contribution to the linewidth from the contribution of the sample's inhomogeneity using standard VNA-FMR modeling \cite{devolder_damping_2013}. This is done by plotting the half FMR linewidth $\Delta f /2$ versus FMR frequency $f_\textrm{FMR}$ [see one example in Fig.~\ref{fmr}(d)]. The Gilbert damping is the curve slope and the line broadening arising from the inhomogeneity of the effective field within the free layer is the zero frequency intercept $\frac{1}{2\gamma_0} \Delta f |_{f=0}$) of the curve. 

%%%%%%%%%%%%%%%%%%%%%%%%%%%%%%%%%%%%%%%%%%%%%%%%%%%%%%%%%%%%%%
\subsection{Boron content and Gilbert damping upon annealing of single MgO free layers} 
\label{BC}%%%%%%%%%%%%%%%%%%%%%%%%%%%%%%%%%%%%%%%%%%%%%%%%%%%%%%%%%%%%%%
%%%%%%%%%%%%%%%%%%%%%%%%%%%%%%%%%%%%%%%%%%%%%%%%%%%%%%%%%%%%%%
Designing advanced free layer in STT-MRAM stacks requires to minimize the Gilbert damping of the used raw material. In Ta/FeCoB/MgO 'single MgO' free layers made of amorphous FeCoB alloys or made of FeCoB that has been just crystallized, a damping of 0.008 to 0.011 can be found typically \cite{devolder_damping_2013, devolder_time-resolved_2016}. (Note that lower values can be obtained but for thicknesses and anisotropies that are not adequate for spin-torque application \cite{liu_ferromagnetic_2011}). The damping of Ta/FeCoB/MgO systems generally degrades substantially when further annealing the already crystallized state \cite{bilzer_study_2006}. Let us emphasize than even in the best cases \cite{liu_ferromagnetic_2011}, the damping of FeCoB based free layers are still very substantially above the values of 0.002 or slightly less than can be obtained on FeCo of Fe bcc perfect single crystals \cite{oogane_magnetic_2006, devolder_compositional_2013}. 

\begin{figure}
\centering
\includegraphics[width=8.3 cm]{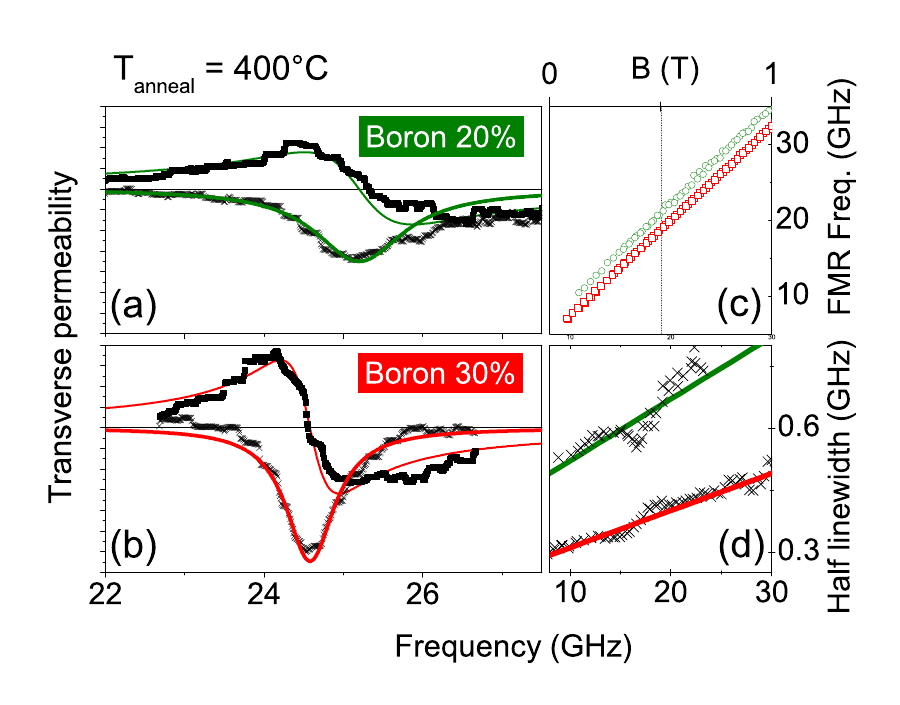}
\caption{(Color online). Properties of single MgO free layers after annealing at 400$^\circ$C. (a) and (b): Real part (narrow lines) and imaginary part (bold lines) of the free layer permeability in a field of 0.7 T. The lines are macrospin fits.  (c) Ferromagnetic resonance frequency versus field curves. (d) Half linewidth versus FMR frequencies. The lines have slopes of $\alpha = 0.009$ (red, B30\%) and $\alpha = 0.015$ (green, B20\%)}
%AE03362_D03 (single MgO POR, like in PRB 2016 and APL 2016) D03 is 14A (Fe60Co20B20%) D09 is 16A, (B30%) It is not thickness matched, but ~ FeCo atoms matched (60+20)/(52.5+17.5) = 16/14.
\label{setup}
\label{BoronContent}
\end{figure}

There are thus potentially opportunities to improve the damping of free layers by material engineering. We illustrate this in fig.~\ref{BoronContent} in which we show that a simple increase of the Boron content is efficient to maintain the damping unaffected, even upon annealing at 400$^\circ$C in a single MgO free layer. Indeed starting from  Ta/FeCoB/MgO 'single MgO' free layers sharing the same damping of 0.009 after annealing at 300$^\circ$C  (not shown), an additional 100$^\circ$C yields $\alpha=0.015$ for the free layer with 20\% of boron, while the boron 30\% free layers keep a damping of $\alpha=0.009$ [see fig.~\ref{BoronContent}(d)]. Meanwhile the anisotropies of these two free layers remain perpendicular [fig.~\ref{BoronContent}(c)] with $\mu_0(H_k-M_s)$ being 0.27 and 0.17 T, respectively, after annealing at 400$^\circ$C. Let us comment on this difference of damping. 

Two mechanisms can yield to extra damping: spin-pumping \cite{tserkovnyak_spin_2002} and spin-flip impurity scattering of the conduction electrons by a spin-orbit process \cite{kambersky_Spin-orbital_2007}. %Spin pumping
Tantalum is known to be a poor spin-sink material as this early transition metal has practically no \textit{d} electrons and therefore its spin-pumping contribution to the damping of an adjacent magnetic layer is weak \cite{tserkovnyak_spin_2002}. We expect a spin pumping contribution to the damping of Ta (2 nm) / FeCoB (1.4 nm) / MgO 'single MgO' free layers that compares with for instance that measured by Mizukami \textit{et al.} on Ta (3 nm) / Fe$_{20}$Ni$_{80}$ (3 nm) which was undetectable \cite{Mizukami_ferromagnetic_2001} since below 0.0001; we therefore expect that the spin-pumping contribution to the total free layer damping is too negligible to account for the differences observed between a free layer and the corresponding perfect single crystals. The main remaining contribution to the damping is the magnon scattering by the paramagnetic impurities within the FeCoB material \cite{rantscher_effect_2007}. %damping and impurities
Indeed the Ta atoms within an FeCoB layer are paramagnetic impurities that contribute to the damping according to their concentration like any paramagnetic dopant; however the effect with Ta is particularly large \cite{devolder_irradiation-induced_2013} as Fe and Co atoms in direct contact with Ta atoms loose part of their moment and get an extra paramagnetic character, an effect usually referred as a "magnetically dead layer". Qualitatively, the Ta atoms in the inner structure of the free layer degrade its damping. 

As the cap of Ta / FeCoB / MgO 'single MgO' free layers contain many Ta atoms available for intermixing, a strong degradation of the damping can be obtained in single MgO systems when interdiffusion occurs. To prevent interdiffusion, we used the following strategy. Amorphous materials (including the glassy metals like FeCoB) are known to be efficient diffusion barriers, as they exhibit atom mobilities that are much smaller than their crystalline counterparts. To avoid the diffusion of Ta atoms to the inner part of the FeCoB free layer, a straightforward way is to maintain the FeCoB in an amorphous state as long as possible during the annealing. In metal-metalloid glasses, the crystallization temperature increases with the metalloid content. \textcolor{black}{In our FeCoB free layers, we find crystallization temperatures of 200, 300, 340 and 375$^\circ$C for boron contents of respectively 10\%, 20\%, 25\% and 30\%}. Increasing the boron content in FeCo alloys is a way to conveniently increase the crystallization temperature and thus preserve a low damping. However since to obtain large TMR requires the FeCoB to be crystalline \cite{yuasa_giant_2004, parkin_giant_2004}, one should engineer the boron content such that the crystallization temperature matches with that used in the CMOS final BEOL annealing of 400$^\circ$C. In practice, we have found that this situation is better approached with a boron content of 30\% than 0\% to 25\%.

%%%%%%%%%%%%%%%%%%%%%%%%%%%%%%%%%%%%%%%%%%%%%%%%%%%%%%%%%%%%%%
%	Figure
%%%%%%%%%%%%%%%%%%%%%%%%%%%%%%%%%%%%%%%%%%%%%%%%%%%%%%%%%%%%%%

%%%%%%%%%%%%%%%%%%%%%%%%%%%%%%%%%%%%%%%%%%%%%%%%%%%%%%%%%%%%%%
\subsection{Gilbert damping in single MgO and dual MgO free layers} %%%%%%%%%%%%%%%%%%%%%%%%%%%%%
\label{SD}%%%%%%%%%%%%%%%%%%%%%%%%%%%%%
In our search to further improve the free layers for STT-MRAM applications, we have compared the damping of optimized 'single MgO' and optimized 'dual MgO' free layers. For a fair comparison, we first compare samples made from FeCoB with the same boron content of 20\% and the same 300$^\circ$C annealing treatement. From Fig.~\ref{fmr}(b) and (c), there is a striking improvement of the FMR linewidths when passing from a single MgO to a dual MgO free layer. To discuss this difference in linewidth, we have separated the Gilbert damping contribution to the linewidth from the contribution of the sample's inhomogeneity. We find that dual MgO systems have systematically a substantially lower damping than single MgO free layers \textcolor{black}{which confirms the trends independently observed by other authors \cite{konoto_effect_2013}}. Damping values as low as low as 0.0039 $\pm 0.005$ were obtained in Ta 3\r{A}-spacer dual MgO stacks [Fig.~\ref{fmr}(d)] \textcolor{black}{after 300$^\circ$C annealing}. Samples with a thicker Ta spacer exhibit an increased damping (not shown).
\textcolor{black}{This trend --lower damping in dual MgO systems --is maintained after 400 $^\circ$C annealing; for that annealing temperature, the best damping are obtained for a slightly different internal configuration of the dual MgO free layer. Indeed a damping of 0.0048 was obtained (not shown) in MgO / Fe$_{52.5}$Co$_{17.5}$B$_{30}$ (1.4 nm) / Ta (0.2 nm) / Fe$_{52.5}$Co$_{17.5}$B$_{30}$ (0.8 nm). This should be compared with that the corresponding single MgO free layer which had a damping of 0.009 for the same annealing condition [Fig. 2(d)].}
This finding is consistent with the results obtained on the single MgO free layer if we assume that the Ta impurities within an FeCoB layer contribute to the damping according to their concentration. Somehow, the number of Tantalum atoms in the initial structure of the free layer sets an upper bound for the maximum degradation of the damping upon its interdiffusion that can occur during the annealing. Notably, the single MgO free layers contain much more Ta atoms (i.e. 2 nm compared to 0.2 to 0.4 nm) available for intermixing: not only the initial number of Ta impurities within the FeCoB layer directly after deposition is larger in the case of single MgO free layer, but in addition a much stronger degradation of the damping can be obtained in single MgO systems when interdiffusion occurs, in line with our experimental findings. This interpretation -- the dominant source of damping is the Ta content -- is further strengthened by the fact that the thickness of the Ta spacer strongly impacts the damping in dual MgO free layers.

Let us now study the spin-torque induced switching process in nanopillars processed from optimized MTJs.
%%%%%%%%%%%%%%%%%%%%%%%%%%%%%%%%%%%%%%%%%%%%%%%%%%%%%%%%%%%%%%
\section{Spin-torque induced switching process} \label{methods}

%%%%%%%%%%%%%%%%%%%%%%%%%%%%%%%%%%%%%%%%%%%%
\subsection{Sample and methods for the switching experiments} \label{methods}
In this section we use two kinds of perpendicularly magnetized MTJ: a 'single MgO' and a 'dual MgO' free layer whose properties are detailed respectively in ref. \cite{devolder_time-resolved_2016} and \cite{devolder_size_2016}. \textcolor{black}{Note that the devices are made from stacks that do not include all the latest material improvement described in the previous sections and underwent only moderate annealing processes of 300$^\circ$C. The 'single MgO' free layer samples include a 1.4 nm FeCoB$_{20\%}$ free layer and a Co/Pt based reference synthetic antiferromagnet. Its most significant properties include \cite{devolder_time-resolved_2016} an areal moment of $M_s t \approx 1.54~\textrm{mA}$, a damping of 0.01, an effective anisotropy field of 0.38 T, a TMR of 150\% %and a resistance-area product is $\textrm{RA}=6.5~ \Omega.\mu \textrm{m}^2$
. The 'dual MgO' devices are made from tunnel junctions with a 2.2 nm thick FeCoB-based free layer and a hard reference system also based on a well compensated \textcolor{black}{[Co/Pt]-based} synthetic antiferromagnet. The perpendicular anisotropy of the (much thicker) free layer is ensured by a dual MgO encapsulation and an iron-rich composition. After annealing, the free layer has an areal moment of $M_s t \approx 1.8~\textrm{mA}$ and an effective perpendicular anisotropy field 0.33 T. Before pattering, standard ferromagnetic resonance measurements indicated a Gilbert damping parameter of the free layer being $\alpha = 0.008$. Depending on the size of the patterned device, the tunnel magnetoresistance (TMR) is 220 to 250\%%, for a stack resistance-area product is $\textrm{RA}=12~ \Omega.\mu \textrm{m}^2$
. }

Both types of MTJs were etched into pillars of various size and shapes, including circles from sub-50 nm diameters to 250 nm and elongated rectangles with aspect ratio of 2 and footprint up to 150 $\times$ 300 nm. The MTJs are inserted in series between coplanar electrodes [Fig.~\ref{setup}(a)] using a device integration scheme that minimizes the parasitic parallel capacitance so as to ensure an electrical bandwidth in the GHz range. The junction properties \cite{devolder_time-resolved_2016, devolder_size_2016} are such that the quasi-static switching thresholds are typically 500 mV%or equivalently GGGGGGG $j_{dc} \approx 4\times 10^{10}~\textrm{A/m}^2$
. Spin-wave spectroscopy experiments similar to ref. \cite{devolder_exchange_2016} indicated that the main difference between the two sample series lies in the FL intralayer exchange stiffness. It is $A=8-9~\textrm{pJ/m}$ in the 2.2 nm thick dual MgO free layers of the samples of ref. \cite{devolder_size_2016} and more usual ($\approx 20$ pJ/m) in the 1.4 nm thick 'single MgO' free layers of the samples of ref. \cite{devolder_time-resolved_2016}.

%\subsubsection{Switching exps}
For switching experiments, the sample were characterized in a set-up whose essential features are described in Fig.~\ref{setup}(a): a slow triangular voltage ramp is applied to the sample in series with a $50~\Omega$ oscilloscope. As the device impedance is much larger than the input impedance of the oscilloscope, we can consider that the switching happens at an applied voltage that is constant during the switching. We capture the electrical signature of magnetization switching by measuring the current delivered to the input of the oscilloscope [Fig.~\ref{setup}(b)]. When averaging several switching events [as conducted in Fig.~\ref{setup}(b)], the stochasticity of the switching voltage induces some rounding of the electrical signature of the transition.  However, the single shot switching events can also be captured (Fig.~\ref{pap250nm}-\ref{sizes}). In that case, we define the time origins in the switching as the time at which a perceivable change of the resistance suddenly happens \textcolor{black}{(see the convention in Fig.~5)}. This will be referred hereafter as the "nucleation" instant. This measurement procedure -- slow voltage ramp and time-resolved current -- entails that the studied reversal regime is the sub-threshold thermally activated reversal switching. \textcolor{black}{This sub-threshold thermally-activated switching regime is not directly relevant to understand the switching dynamics in memory devices in which the switching will be forced by short pulses of substantially higher voltage \cite{hahn_time-resolved_2016}. However elucidating the sub-threshold switching dynamics is of direct interest for the quantitative understanding of read disturb errors that may happen at applied voltages much below the writing pulses}. Note finally that sending directly the current to the oscilloscope has a drawback: the current decreases as the MTJ area such that the signal-to-noise ratio of our measurement degrades substantially for small device areas (Fig.~\ref{sizes}). As a result, the comfortable signal-to-noise ratio allows for a very precise determination of the onset of the reversal in large devices, but the precision degrades substantially to circa 500 ps for the smallest (40 nm) investigated devices.

%%%%%%%%%%%%%%%%%%%%%%%%%%%%%%%%%%%%%%%%%%%%%%%%%%%%%%%%%%%%%%
%	Figure
%%%%%%%%%%%%%%%%%%%%%%%%%%%%%%%%%%%%%%%%%%%%%%%%%%%%%%%%%%%%%%
\begin{figure}
\centering
\includegraphics[width=9.3 cm]{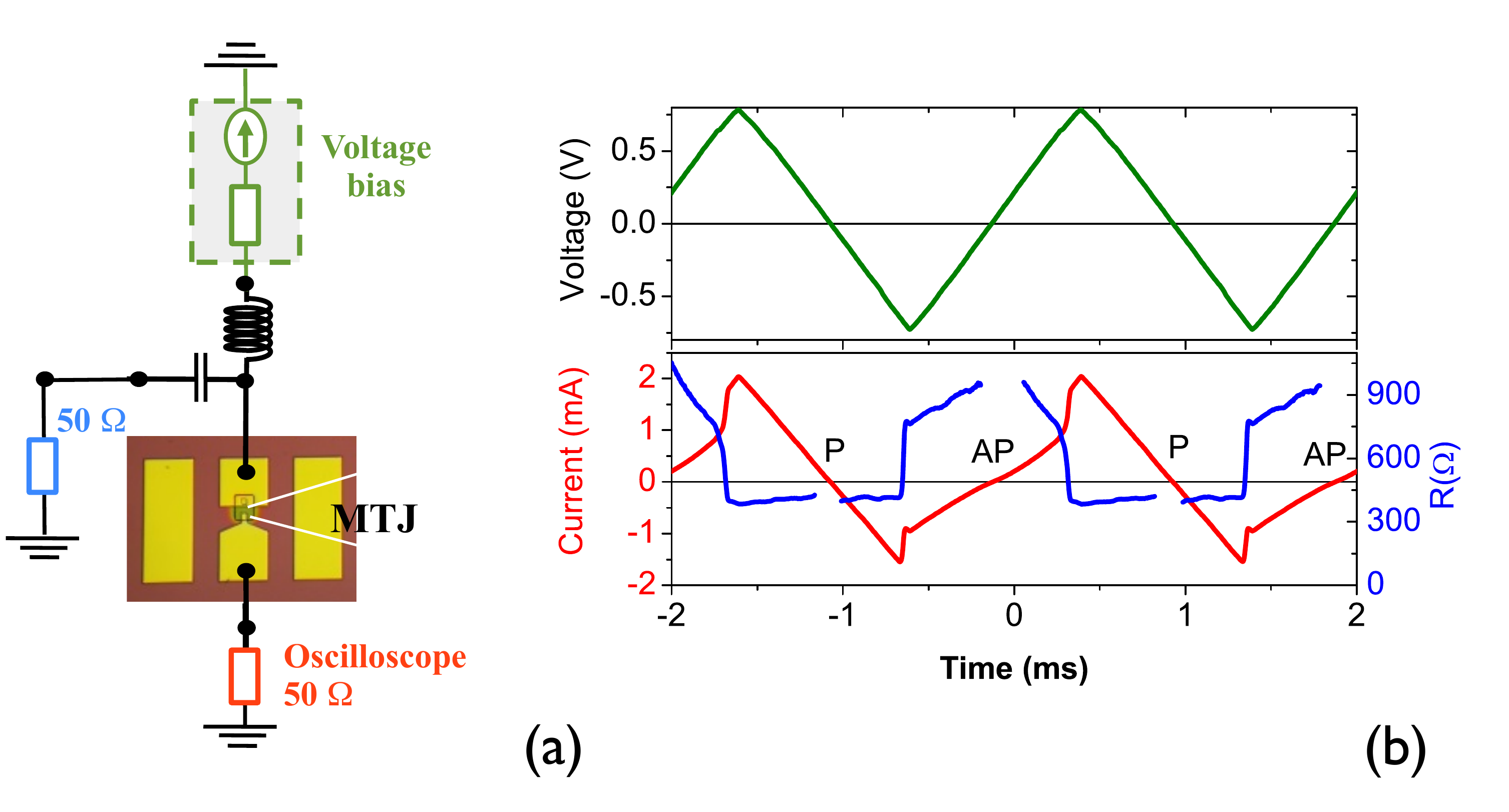}
\caption{(Color online). (a) Sketch of the experimental set-up. Measurement procedure: the device is biased with a triangular kHz-rate voltage (green) and the current (red) is monitored by a fast oscilloscope connected in series. (b) The switching transitions are seen as abrupt changes of the current (red) followed by a change of the current slope. The resistance (blue) can be computed from the voltage-to-current ratio when the current is sufficiently non-zero. In this figure, the displayed currents and resistances are the averages over 1000 events for a 250 nm device with a dual MgO free layer of thickness 2.2 nm and a weak exchange stiffness.}
\label{setup}
\end{figure}

\subsection{Switching results} \label{results}
\begin{figure}
\centering
\includegraphics[width=8 cm]{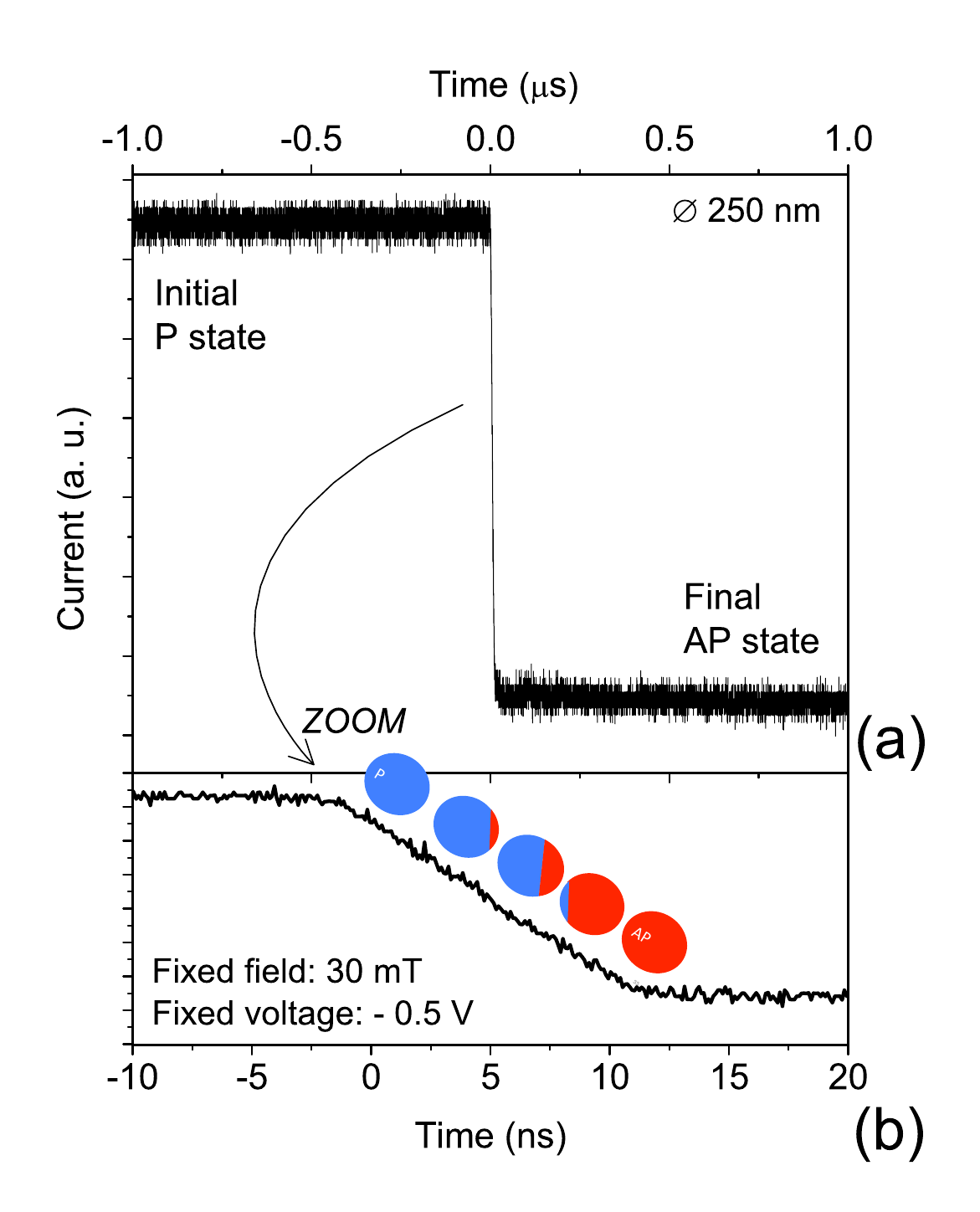}
\caption{(Color online). Single-shot time-resolved absolute value of the current during a spin-torque induced switching for parallel to antiparallel switching for a circular device of diameter 250 nm made with a weak exchange stiffness, dual MgO 2.2 nm thick free layer. (a) Two microsecond long time trace, illustrating that the switching is complete, free of back-hopping phenomena, and occurs between two microwave quiet states. (b) 30 ns long time trace illustrating the regular monotonic change of the device conductance during the switching.}
\label{pap250nm}
\end{figure}

\begin{figure}
\centering
\includegraphics[width=9 cm]{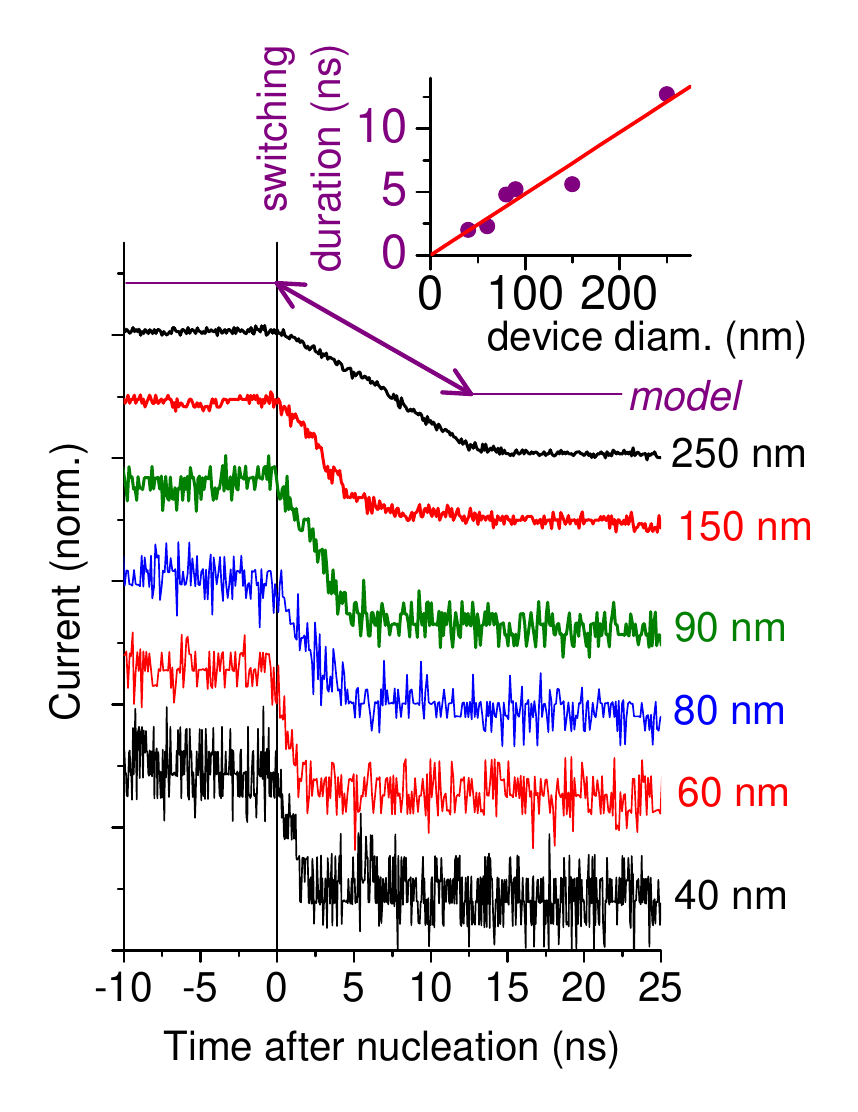}
\caption{(Color online). Single-shot time-resolved conductance traces for parallel to antiparallel switching events occurring at at -0.5 V for circular devices of various diameters. The curves are for the devices whose dual MgO free layer has a thickness of 2.2 nm and has a weak exchange stiffness. The curves have been vertically offset and vertically normalized to ease the comparison. The time origins and switching durations are chosen at the perceivable onset and end of the conductance change\textcolor{black}{: they are defined by fitting the experimental conductance traces by 3 segments (see the sketch labelled "model")}. Inset: duration of the switching events versus free layer diameter (symbols) and linear fit thereof with an inverse slope of 20 m/s.}
\label{sizes}
\end{figure}

In samples whose (i) reference layers are sufficiently fixed to ensure the absence of back-hopping \cite{devolder_time-resolved_2016} and (ii) in which the stray field from the reference layer is rather uniform \cite{devolder_size_2016}, optimized compensation of the stray field of the reference layers leads to a STT-induced switching with a simple and abrupt electrical signature [Fig.~\ref{pap250nm}(a)]. If examined with a better time resolution, the switching event [Fig.~\ref{pap250nm}(b)] appears to induce a monotonic ramp-like evolution of the device conductance. For a given MTJ stack, the switching voltage is practically independent from the device size and shape in our interval of investigated sizes (not shown). This finding is consistent with the consensual conclusion that the switching energy barrier is almost independent from the device area \cite{chaves-oflynn_thermal_2015, sun_spin-torque_2013} for device areas above 50 nm.  In spite of this quasi-independence of the switching voltage and the device size, the switching duration was found to strongly depend on device size (Fig.~\ref{sizes}); we have found that smaller devices switch faster, and the trend is that the switching duration correlates linearly with the longest dimension of the device. This is shown in Fig.~\ref{sizes}: 40 nm devices switch in typically 2 to 3 ns whereas devices that are 6 times larger switch in 10 to 15 ns. 

Such a reversal path can be interpreted this way: once a domain is nucleated at one edge of the device, the domain wall sweeps irreversibly through the system at a velocity set by the applied voltage [sketch in Fig.~\ref{pap250nm}(b)]. The average domain wall speed is then about 20 nm/ns for the low-exchange-free-layers of ref. \cite{devolder_size_2016}. The other devices (not shown but described in ref. \cite{devolder_time-resolved_2016}) based on a 'single MgO' free layer with a more bulk-like exchange switch with a substantially higher apparent domain wall velocity, reaching 40 m/s.

%%%%%%%%%%%%%%%%%%%%%%%%%%%%%%%%%%%%%%%%%%%%%%%%%%%%%%%%%%%%%%
%	Figure
%%%%%%%%%%%%%%%%%%%%%%%%%%%%%%%%%%%%%%%%%%%%%%%%%%%%%%%%%%%%%%

%%%%%%%%%%%%%%%%%%%%%%%%%%%%%%%%%%%%%%%%%%%%%%%%%%%%%%%%%%%%%%
%	Figure
%%%%%%%%%%%%%%%%%%%%%%%%%%%%%%%%%%%%%%%%%%%%%%%%%%%%%%%%%%%%%%

%%%%%%%%%%%%%%%%%%%%%%%%%%%%%%%%%%%%
 \subsection{Switching Model: domain wall-based dynamics} \label{model}
 
To model the switching, we assume that there is a domain wall (DW) which lies at a position $q$ and moves along the longest axis $x$ of the device. The domain wall is assumed to be straight along the $y$ direction, as sketched in Fig.~\ref{pap250nm}(b). We describe the wall in the so-called 1D model \cite{thiaville_micromagnetic_2005}: the wall is assumed to be a rigid object of fixed width $\pi \delta$ presenting a tilt $\phi$ of its magnetization in the device plane; by convention $\phi=0$ is for a wall magnetization along $x$, i.e. a N\'eel wall. 

The local current density at the domain wall position is written $j$.  The wall is subjected to %both an in-plane field $H_x$ and 
an out-of-plane field $H_z$ assumed to vary slowly in space at the scale of the DW width. $j$ is assumed to transfer $p \approx 1$ spin per electron to the DW by a pure Slonczewski-like STT. We define 
\begin{equation} \label{sigma}
\sigma = \frac{\hbar}{2e} \frac{\gamma_0}{\mu_0 M_S t}
\end{equation}

as the spin-transfer efficiency in unit such that $\sigma j $ is a frequency.
With typical FeCoB parameters, i.e. magnetization $M_s \in [1.1,~1.4]$ MA/m and free layer thickness $t \in [1.4,~2.2]$ nm, we have $\sigma \in [0.018,~0.036]$ Hz / (A/m$^2$) where the lowest value corresponds to the largest areal moment $M_s t$. With switching current density of the order of $4\times 10^{10}~\textrm{A/m}^2$, this yields  $\sigma j_{dc}$ between 0.72 and 1.4 GHz.

Following ref. \cite{cucchiara_domain_2012}, the wall position $q$ and and wall tilt $\phi$ are linked by the two differential equations: 

\begin{equation}
 \dot{\phi} + \frac {\alpha}{\delta} \dot{q} = \gamma_0 H_z ~,
  \label{1Dmodela} 
  \end{equation}
  
  \begin{equation}
  \frac {\dot{q}}{\delta} - \alpha \dot{\phi} 
 = \sigma j_\textrm{dc} + \frac{\gamma_0 H_{\textrm{DW}} }{2} \sin(2 \phi) %- {\frac{\pi}{2}}\gamma_0 H_x \sin \phi
  \label{1Dmodelb} 
 \end{equation}
  
in which $\pi \delta$ is the width of a Bloch domain wall in an ultrathin film, with $\delta^2 = 2A/(\mu_0 M_s H_k^\textrm{eff})$ where $A$ is the exchange stiffness. A wall parameter $\delta=12~\textrm{nm}$ will be assumed for the normal exchange 1.4 nm free layer from various estimates including ref. \cite{devolder_exchange_2016} for the  exchange stiffness and ref. \cite{devolder_evolution_2016} for the anisotropy of the free layer. The domain wall stiffness field \cite{mougin_domain_2007} $H_{\textrm{DW}}$ is the in-plane field that one would need to apply to have the wall transformed from a Bloch wall to a N\'eel wall. As it expresses the in-plane demagnetization field within the wall, it depends on the wall width $\pi \delta$ and on the wall length when the finite size of the device constrains the wall dimensions.

Using \cite{mougin_domain_2007}, the domain wall stiffness field can be estimated to be at the most 20 mT in our devices. In circular devices, the domain wall has to elongate upon its propagation \cite{chaves-oflynn_thermal_2015} such that the domain wall stiffness field $H_{\textrm{DW}}$ depends in principle on the DW position. It should be maximal when the wall is along the diameter of the free layer. However we will see that  $H_{\textrm{DW}}$ is not the main determinant of the dynamics. 
Indeed in the absence of stray field and current, the Walker field $H_{\textrm{Walker}}$ is proportional to the domain wall stiffness field times the damping parameter, i.e.  $H_{\textrm{Walker}} = \alpha H_{\textrm{DW}} /2$. As the samples required for STT switching are typically made of low damped materials with $\alpha < 0.01$, the Walker field is very small and likely to be smaller than the stray fields emanating from either the reference layers or the applied field. This very small Walker field has implications: in practice as soon as there is some field of some applied current, any domain wall in the free layer is bound to move in the Walker regime and to make the back-and-forth oscillatory movements that are inherent to this regime. The DW oscillates at a generally fast (GHz) frequency \cite{thiaville_domain-wall_2006} such that only the time-averaged velocity matters to define how much it effectively advances.
 
 %%%%%%%%%%%%%%%%%%%%%%%%%%%%%%%%%%%%

To see quantitatively the effect of a constant current on the domain wall dynamics, we assume that the sample is invariant along the domain wall propagation direction (x) (like in an hypothetical stripe-shaped sample). Solving numerically Eq.~\ref{1Dmodela} and \ref{1Dmodelb}, we find that the Walker regime is maintained for $j_\textrm{dc} \neq 0$ (not shown). Two points are worth noticing: \\
The time-averaged domain wall velocity $ \langle \dot{q} \rangle$ varies linearly with the applied current density.  When in the Walker regime, the current effect can be understood from Eq.~\ref{1Dmodelb}. Indeed the $\sin(2\phi)$ term essentially averages out in a time integration as $\phi$ is periodic, and the term $\alpha \dot{\phi}$ is neligible, such that the time-averaged wall velocity reduces to: 
\begin{equation}
 \langle \dot{q} \rangle \approx   \delta \sigma j_\textrm{dc} \label{averagevelocity}
 \end{equation}
For $\delta=12$~nm and $\sigma j$ in the range of 1.4 GHz at the switching voltage for the bulk-like exchange stiffness sample with free layer thcikness 1.4 nm, the previous equation would predict a time-averaged domain wall velocity of 17 m/s (or nm/ns) during the switching.  More compact domain walls are expected for the samples with a weaker exchange stiffness; the twice lower $\sigma j \approx 0.72$ GHz related to the larger thickness would reinforce this trend to a much a lower domain wall velocity (9 m/s for our material parameters estimates).  This expectation compares qualitatively well with our experimental findings of slower walls in weakly exchanged materials (Fig.~\ref{sizes}). 

We wish to emphasize that Eq.~\ref{averagevelocity} can be misleading regarding the role of damping. Indeed a too quick look at Eq.~\ref{averagevelocity} could let people wrongly conclude the domain wall velocity is essentially set by the areal moment $M_s t$ and that the wall velocity under STT from a current perpendicular to the plane (CPP current) is independent from the damping factor (see Eq.~\ref{sigma}). However this is not the case as the switching current $j_\textrm{dc}$ is a sweep-rate-dependent and temperature-determined fraction $\eta \in [\frac{1}{2}, 1]$ of the zero temperature instability current $j_{c0}$ of a macrospin in the parallel state, which reads \cite{sun_spin-current_2000, sun_magnetoresistance_2008}: 

\begin{equation}
j_{c0} = \alpha \frac{4e}{\hbar}  \frac{1+p^2}{p} \frac{\mu_0 M_s t H_{k}^\textrm{eff}}{2} \label{jc0}
\end{equation} where $p \approx 1$ is an effective spin polarization.

Using Eq.~\ref{sigma},~\ref{averagevelocity} and \ref{jc0}, the time-averaged wall velocity at the practical switching voltage is:
 \begin{equation}
 \langle \dot{q} \rangle \approx  \alpha ~ \delta ~ \gamma_0 H_{k}^\textrm{eff} ~\eta
 \end{equation}

This expression indicates that the samples performing best in term of switching current (minimal damping and easy nucleation thanks to a small exchange) will host domain walls that are inherently slow when pushed by the CPP current in the Walker regime. The domain wall speed scales with the domain wall width, which may be the reason why the low exchange stiffness samples host domain walls that are experimentally slower.

To summarize, once nucleated at the instability of the uniformly magnetized state at $j_\textrm{dc} = \eta j_{c0}$, the domain wall flows in a Walker regime through the device. The switching duration varies thus simply with the inverse current: 

\begin{equation}
\tau_{\textrm{switch}} =  \frac{L}{\delta \sigma j_\textrm{dc}} ~\approx ~\frac{L}{\delta} ~\times ~\frac{1}{\alpha  \gamma_0 H_{k}^\textrm{eff}} ~\times ~\frac{1} {\eta}
\label{tauswitch} 
\end{equation}

Let us comment on this equation which is the main conclusion of this section. The underlying simplifications are: (i) a rigid wall (ii) that does not sense the sample's edges (iii) that moves at a speed equal to its average velocity in the Walker regime (iv) at a switching voltage that is independent from the sample geometry. Under these assumptions, the duration of the switching scales with the length $L$ of the sample, as observed experimentally. It also scales with the inverse of the zero-field ferromagnetic resonance linewith $2 \alpha  \gamma_0 H_{k}^\textrm{eff}$. The practical switching voltage is below the zero temperature macrospin switching voltage by a factor $\eta$, which gathers the effect of the thermal activation and of the sweeping rate of the applied voltage \cite{koch_time-resolved_2004}. $\eta \approx 1/2$ for quasi-static experiments like reported here and $\eta \rightarrow 1$ for experiments in which the voltage rise time $V_\textrm{max}/ \dot{V}$ is short enough compared to the switching duration (Eq.~\ref{tauswitch}). 

\section{Summary and conclusion}
In summary, we have investigated the Gilbert damping of advanced free layer designs: they comprise FeCoB alloys with variable B contents from 20 to 30\% and are organized in the single MgO or dual MgO free layer configuration fully embedded in functional STT-MRAM magnetic tunnel junctions. Increasing the boron content increases the cristallization temperature, thereby postponing the onset of elemental diffusion within the free layer. This reduction of the interdiffusion of the Ta atoms helps maintaining the Gilbert damping at a low level without any penalty on the anisotropy and the transport properties. Thereby, increasing the Boron content to at least 30\% is beneficial for the thermal robustness of the MTJ up to the 400$^\circ$ required in CMOS back-end of line processing. In addition, we have shown that dual MgO free layers have a substantially lower damping than their single MgO counterparts, and that the damping increases as the thickness of the Ta spacer within dual MgO free layers. This indicates that the dominant source of extra damping is the presence of Ta impurities within the FeCoB alloy. 
Using optimized MTJs, we have studied the duration of the switching events as induced by spin-transfer-torque. Our experimental procedure -- time-resolving the switching with a high bandwidth but during slow voltage sweep -- ensures that we are investigating only sub-threshold thermally activated switching events. In optimal conditions, the switching induces a ramp-like monotonic evolution of the device conductance that we interpret as the sweeping of a domain wall through the device. The switching duration is roughly proportional to the device size: the smaller the device, the faster it switches. We studied two MTJ stacks and found domain wall velocities from 20 to 40 m/s.  
A simple analytical model using a rigid wall approximation can account for our main experimental findings. The domain wall velocity is predicted to scale linearly with the current for device sizes much larger than the domain wall widths. The domain wall velocity depends on the material parameters, such that the samples with the thinnest domain walls will be the ones that host the domain walls with the lowest mobilities. Schematically, material optimization for low current STT-induced switching (i.e. in practice: fast nucleation because of low exchange stiffness $A$ and low damping $\alpha$) will come together with slow STT-induced domain wall motion at least in the range of device sizes in which the STT-induced reversal proceeds through domain wall motion. \textcolor{black}{If working with STT-MRAM memory cells made in the same range of device sizes, read disturb should be minimal (if not absent) provided that the voltage pulse used to read the free layer magnetization state has a duration much shorter than the time needed for a domain wall to sweep through the device at that voltage (Eq.~\ref{tauswitch}).}

\section*{Acknowledgment}

This work is supported in part by IMEC's Industrial Affiliation Program on STT-MRAM device, in part by the Samsung Global MRAM Innovation Program and in part by a public grant overseen by the French National Research Agency (ANR) as part of the ÒInvestissements dÕAvenirÓ program (Labex NanoSaclay, reference: ANR-10-LABX-0035). T. D. would like to thank Andr\'e Thiaville, Paul Bouquin and Felipe Garcia-Sanchez for useful discussions.

%\bibliography{bib.bib}
\bibliographystyle{IEEEtran}


\end{document}